\documentclass[twocolumn,aps,prd,10pt,superscriptaddress,longbibliography,floatfix,citeautoscript]{revtex4-2}

\usepackage{amsmath}
\usepackage{amssymb}
\usepackage{bm}
\usepackage{dcolumn}
\usepackage{glossaries}
\usepackage{graphicx}
\usepackage{multirow}
\usepackage{fancyvrb}
\usepackage[linesnumbered,ruled]{algorithm2e}
\usepackage{siunitx}

\usepackage[usenames,dvipsnames,svgnames]{xcolor}
\usepackage{hyperref}
\hypersetup{
    pdfnewwindow=true,      % links in new window
    colorlinks=true,        % false: boxed links; true: colored links
    linkcolor=Blue,         % color of internal links
    citecolor=Blue,         % color of links to bibliography
    filecolor=Blue,         % color of file links
    urlcolor=Blue           % color of external links
}

\usepackage{listings}	
\setacronymstyle{long-short}
\newacronym{dft}{DFT}{density functional theory}
\newacronym{gpu}{GPU}{graphics processing units}
\newacronym{pbe}{PBE}{Perdew-Burke-Ernzerhof}
\newacronym{md}{MD}{molecular dynamics}
\newacronym{mbd}{MBD}{many-body dispersion}
\newacronym{mlp}{MLP}{machine-learned potential}
\newacronym{nep}{NEP}{neuroevolution potential}
\newacronym{rmse}{RMSE}{root mean square error}
\newacronym{snes}{SNES}{separable natural evolution strategy}
\newacronym{vdw}{vdW}{van-der-Waals}
\newacronym{sp}{SP}{saddle point}
\newacronym{2d}{2D}{two-dimensional}
\newacronym{hnemd}{HNEMD}{homogeneous non-equilibrium molecular dynamics}
\newacronym{mof}{MOF}{metal-organic framework}

\DeclareSIUnit\angstrom{\text{Å}}
\DeclareSIUnit{\atom}{atom}
\DeclareSIUnit{\step}{step}
\DeclareSIUnit{\atomstepsecond}{\atom\step\per\second}

\newcolumntype{d}{D{.}{.}{-1}}

\begin{document}

\title{Combining the D3 dispersion correction with the neuroevolution machine-learned potential}

\author{Penghua Ying}
\email{hityingph@163.com}
\affiliation{Department of Physical Chemistry, School of Chemistry, Tel Aviv University, Tel Aviv, 6997801, Israel}

\author{Zheyong Fan}
\email{brucenju@gmail.com}
\affiliation{College of Physical Science and Technology, Bohai University, Jinzhou 121013, P. R. China}

\date{\today}

\begin{abstract}
Machine-learned potentials (MLPs) have become a popular approach of modelling interatomic interactions in atomistic simulations, but to keep the computational cost under control, a relatively short cutoff must be imposed, which put serious restrictions on the capability of the MLPs for modelling relatively long-ranged dispersion interactions. 
In this paper, we propose to combine the neuroevolution potential (NEP) with the popular D3 correction to achieve a unified NEP-D3 model that can simultaneously model relatively short-ranged bonded interactions and  relatively long-ranged dispersion interactions. 
We show the improved descriptions of the binding and sliding energies in bilayer graphene can be obtained by the NEP-D3 approach compared to the pure NEP approach. 
We implement the D3 part into the GPUMD package such that it can be used out of the box for many exchange-correlation functionals. 
As a realistic application, we show that dispersion interactions result in approximately a 10\% reduction in thermal conductivity for three typical metal-organic frameworks.
\end{abstract}

\maketitle

\section{Introduction}

\Gls{mlp} \cite{behler2016jcp} is an emerging approach for modelling the interatomic interactions in materials. To achieve a linear scaling of the computational cost with respect to the system size, a \gls{mlp} must be constructed based on local descriptors \cite{Musil2021cr}. The descriptor for an atom is usually constructed based on the positions of its neighbors within a certain cutoff $R_{\rm c}$. The average number of neighbors for an atom, hence the computational cost of a \gls{mlp}, is proportional to $R_{\rm c}^3$. Therefore, in practice, $R_{\rm c}$ is usually chosen to be a few \SI{}{\angstrom}. This length is usually sufficient for describing the bond interactions in typical materials, but is not sufficient for describing the London dispersion interactions that can extend to one to a few \SI{}{\nano\meter}. However, the dispersion interactions are important in describing e.g., the interlayer attractions in the so-called \gls{vdw} layered materials \cite{Geim2013, mandelli2019princess} and structural transformation between a narrow-pore and large-pore phase in flexible \glspl{mof} \cite{walker2010flexibility, wieme2018tuning}.

To address this challenge, a few attempts have been made to augment \glspl{mlp} with dispersion corrections. Wen and Tadmor \cite{wen2019prb} added an attractive $-C_6/r_{ij}^6$ term multiplied by some switching/damping functions to a \gls{mlp}, where $r_{ij}$ is the distance between atoms $i$ and $j$ and $C_6>0$ is a fitting parameter. There are four additional fitting parameters in the switching functions \cite{wen2019prb}. They have obtained quite good description for the binding and sliding energies in bilayer graphene. By adding a similar $-C_6/r_{ij}^6$ term, Deringer \textit{et al.} \cite{Deringer2020NC} 
and Rowe \textit{et al.} \cite{Rowe2020jcp} constructed general-purpose \glspl{mlp} for phosphorous and carbon systems. Muhli \textit{et al.} \cite{Muhli2021prb} developed a more sophisticated method that determines the dispersion coefficient and damping function using a local descriptor. 

Despite these achievements, it still requires quite a lot efforts to determine the dispersion interactions for a \gls{mlp} model. The determination of the dispersion coefficient and the damping function is species and system dependent, and the process is thus not very systematical and transferable. Indeed, it is known that the dispersion coefficient is environment dependent, which has been taken into account in many popular dispersion corrections to \gls{dft}, such as the D3 approach \cite{Grimme2010jcp,grimme2011effect}. To our best knowledge, the D3 approach has not been combined with \glspl{mlp} to perform large-scale atomitstic simulations. One reason is that there is so far no efficient implementation of D3 for the use with classical potentials. In this work, we make an efficient \gls{gpu} implementation of D3 into the \textsc{gpumd} package \cite{fan2017cpc} and combine it with the machine-learned \gls{nep} \cite{fan2021prb,fan2022jpcm,fan2022jcp} to form a \gls{nep}-D3 approach. This approach inherits all the merits of D3 and applies to the 94 elements H-Pu for a large number of \gls{dft} functionals. Due to the separability of the \gls{nep} and D3 parts in our approach, it also allows for isolating the role of dispersion interactions in affecting specific physical properties. We will use two example systems, bilayer graphene and \glspl{mof}, to demonstrate the convenience, accuracy, efficiency, generality, and usefulness of the \gls{nep}-D3 approach. 

\gls{nep} is a neural-network-based \gls{mlp} which got its name due the use of an evolutionary algorithm for training the free parameters \cite{fan2021prb,fan2022jpcm,fan2022jcp}. In this method, the total energy of a system is taken as the sum of the site energies at each atom, which is taken as the output of a neural-network, as first proposed by Behler and Parrinello \cite{behler2007prl}. The input layer of the neural-network consists of a number of features called descriptor components. The design of descriptor components is crucial for a \gls{mlp} and different \glspl{mlp} presently in use differ from each other mainly by the descriptor. In the very first version of \gls{nep} \cite{fan2021prb}, it has been realized to be beneficial to have two different kinds of descriptor components, the radial ones and the angular ones. The radial descriptor components depend only the interatomic distances, which are relatively cheaper to calculate, while the angular descriptor components also depend on bond angles and are relatively more expensive to calculate. Therefore, it has been suggested to use a relatively longer cutoff for the radial descriptor components $r_{\rm c}^{\rm R}$ in combination with a relatively shorter cutoff for the angular descriptor components $r_{\rm c}^{\rm A}$ when this is beneficial \cite{fan2021prb}. For example, this strategy has been used for modelling a number of \gls{vdw} structures \cite{dong2023ijhmt,eriksson2023tuning} that have both strong covalent bonds and weak vdW interactions, where $ r_{\rm c}^{\rm R} $ was taken to be about \SIrange{7}{8}{\angstrom} and $ r_{\rm c}^{\rm A} $ \SIrange{3}{4}{\angstrom}. While constructing a pure \gls{nep} that incorporates dispersion interactions is feasible, the primary objective of this work is to introduce the \gls{nep}-D3 method that addresses dispersion interactions in a more elegant manner. We will compare results from the two approaches in terms of both accuracy and computational efficiency.

\section{Results and discussion}

\subsection{Implementation of the D3 dispersion correction into GPUMD}

The D3 dispersion correction has contributions from a ``two-body'' (the meaning of the quotes will be made clear soon) part and a three-body part. However, it has been recommended not to include the three-body part \cite{Grimme2010jcp}. We therefore only considered the two-body part, which is also the choice of our reference \gls{dft} code Vienna Ab initio Simulation Package (\textsc{vasp}) \cite{Kresse1996PRB, Kresse1999PRB}. We chose to use the Becke-Johnson damping which does not introduce spurious force at short distances \cite{grimme2011effect}. The total D3 energy can be expressed as
\begin{equation}
U^{\rm D3} =\sum_{i} U_i^{\rm D3};
\end{equation}
\begin{align}
U_i^{\rm D3} 
&= -\frac{1}{2}\sum_{j\neq i} \frac{s_6 C_{6ij}}{r^6_{ij} + (a_1R_0+a_2)^6} \nonumber \\
&-\frac{1}{2} \sum_{j\neq i} \frac{s_8 C_{8ij}}{r^8_{ij} + (a_1R_0+a_2)^8}, 
\end{align}
where $ r_{ij} $ is the distance between atoms $ i $ and $j$, $s_6$, $s_8$, $a_1$, and $a_2$ are parameters depending on the chosen exchange-correlation functional, $C_{6ij}$ and $C_{8ij}$ are the dispersion coefficients for the $ij$ atom pair, and $R_0$ is taken as the geometric mean of tabulated parameters for each sepecies. The summation of $j$ is over the neighbors of $i$ within a cutoff $R_{\rm pot}$. 

The two dispersion coefficients are related by $C_{8ij} = C_{6ij} R_0^2$.
The dispersion coefficient $C_{6ij}$ is calculated as a function of the coordination numbers $n_i$ and $n_j$,
\begin{equation}
C_{6ij}(n_i, n_j) = 
\frac{\sum_a\sum_b C_{6ijab}^{\rm ref} 
e^{-4[(n_{i}-n_{ia}^{\rm ref})^2 
+ (n_{j}-n_{jb}^{\rm ref})^2]}}
{\sum_a\sum_b e^{-4[(n_{i}-n_{ia}^{\rm ref})^2 
+ (n_{j}-n_{jb}^{\rm ref})^2]}},
\end{equation}
where the summations of $a$ and $b$ are over the numbers of reference points for atoms $i$ and $j$, respectively. Here, $n_{ia}^{\rm ref}$ is the $a$-th reference coordination number for atom $i$, $n_{jb}^{\rm ref}$ is the $b$-th reference coordination number for atom $j$,  and $C_{6ijab}^{\rm ref}$ is the $(a,b)$ reference dispersion coefficient for the atom pair $(i,j)$.

The coordination number for atom $i$ is defined as
\begin{equation}
n_i = \sum_{j\neq i} \frac{1}{1 + e^{-16\left[ (R^{\rm cov}_{i} + R^{\rm cov}_{j}) /r_{ij}-1 \right]}},
\end{equation}
where $R^{\rm cov}_{i}$ is the effective covalent radius of atom $i$. The summation of $j$ is over the neighbors of $i$ within a cutoff $R_{\rm cn}$. Because the coordination number of atom $i$ depends on its neighbors, it is clear to see that the ``two-body'' part of D3 is not a truly two-body (pairwise) potential, but a many-body potential. This is the reason for using the quotes. For a many-body potential, the general formulation for force, virial, and heat current has been established before \cite{fan2015prb,Gabourie2021prb}. The force acting on atom $i$ can be written as 
\begin{equation}
\bm{F}^{\rm D3}_{i}  = \sum_{j\neq i} \bm{F}^{\rm D3}_{ij},
\end{equation}
where
\begin{equation}
\bm{F}^{\rm D3}_{ij}  = \frac{\partial U_i^{\rm D3}}{\partial \bm{r}_{ij}} - \frac{\partial U_j^{\rm D3}}{\partial \bm{r}_{ji}},
\end{equation}
and $\bm{r}_{ij} \equiv \bm{r}_{j} - \bm{r}_{i}$.
The per-atom virial can be defined as
\begin{equation}
\mathbf{W}^{\rm D3}_i = \sum_{j\neq i} \bm{r}_{ij} \otimes \frac{\partial U_j^{\rm D3}}{\partial \bm{r}_{ji}}, 
\end{equation}
and the per-atom heat current can be expressed as
\begin{equation}
\bm{J}^{\rm D3}_i = \mathbf{W}^{\rm D3}_i \cdot \bm{v}_i,
\end{equation}
where $\bm{v}_i$ is the velocity of atom $i$.
The efficient \gls{gpu} implementation of D3 then follows the general algorithms for many-body potentials \cite{fan2017cpc}.
From a practical point of view, the use of D3 only requires three inputs: the exchange-correlation functional, the cutoff for the potential $R_{\rm pot}$, and the cutoff for the coordination number $R_{\rm cn}$. The combined \gls{nep}-D3 potential is simply a sum of D3 and \gls{nep} energies, 
\begin{equation}
U^{\rm NEP-D3} = U^{\rm NEP} + U^{\rm D3},
\end{equation}
where $U^{\rm NEP}$ is the NEP energy as detailed in previous works \cite{fan2021prb,fan2022jpcm,fan2022jcp}. The \gls{nep} model here can also be replaced by the \gls{nep}-ZBL model \cite{liu2023prb} where the Ziegler-Biersack-Littmark (ZBL) potential is used to describe the strong repulsive forces at extremely short distances. 

\begin{figure}[ht]
\centering
\includegraphics[width=\columnwidth]{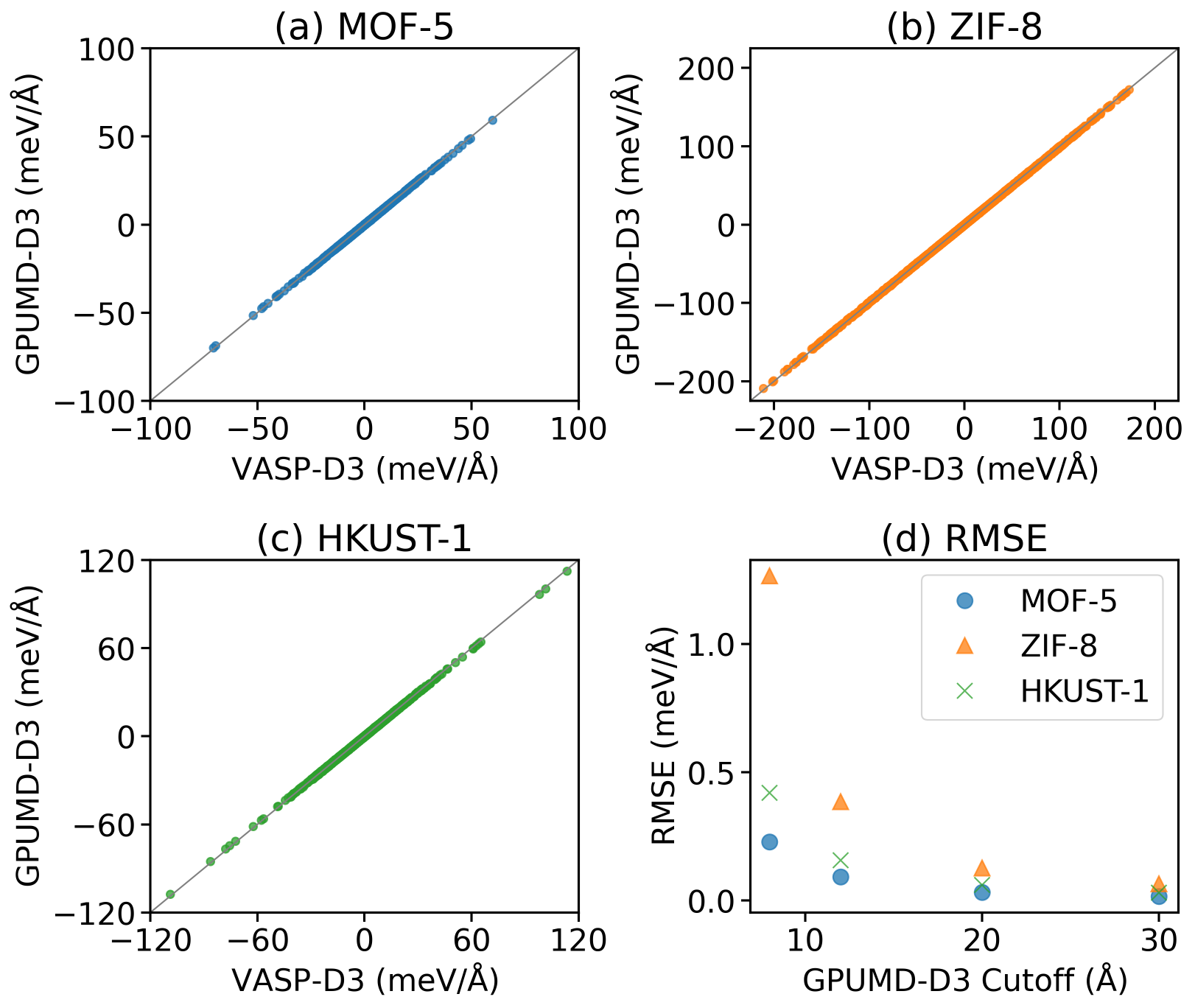}
\caption{(a)-(c) The D3 forces computed from \textsc{gpumd} implementation with $ R_{\rm pot} = \SI{12}{\angstrom}$ and $R_{\rm cn} = \SI{6}{\angstrom}$ as compared to those from \textsc{vasp} implementation with $ R_{\rm pot} = \SI{50.2}{\angstrom}$ and $R_{\rm cn} = \SI{20}{\angstrom}$.  For each \gls{mof}, we generated 10 structures with random cell deformations (from -3 \% to 3\%) and atom displacements (less than \SI{0.1}{\angstrom}), starting from optimized ones. (d) RMSE between \textsc{gpumd} and \textsc{vasp} D3 forces as a function of $R_{\rm pot}$ in \textsc{gpumd}, where $R_{\rm cn}$ is taken as $R_{\rm pot}/2$.
}
\label{fig:crosscheck}
\end{figure}

To confirm the correctness of our GPU implementation of D3 in \textsc{gpumd} and to evaluate the effects of cutoffs, we take three \gls{mof} materials (MOF-5, ZIF-8, and HKUST-1) as studied before using \gls{nep} models \cite{ying2023sub} and compare the calculated forces to those from \textsc{vasp} (using the \verb"IVDW=12" option). The results are shown in \autoref{fig:crosscheck}.
The \textsc{vasp} code uses $R_{\rm pot} = \SI{50.2}{\angstrom}$ and $R_{\rm cn} = \SI{20}{\angstrom}$ as defaults. With $ R_{\rm pot} = \SI{12}{\angstrom}$ and $R_{\rm cn} = \SI{6}{\angstrom}$, the forces calculated from \textsc{gpumd} already agree quite well with those from \textsc{vasp}. With increasing cutoff, the \glspl{rmse} between \textsc{gpumd} and \textsc{vasp} implementations, eventually diminish. While D3 is almost free in \gls{dft} calculations, it can take a considerable portion of time for \glspl{mlp}, which is particularly true for the highly efficient \gls{nep} approach. Therefore, a trike between accuracy and efficiency must be made in selecting the D3 cutoffs in \gls{nep}-D3. We used $R_{\rm pot} = \SI{12}{\angstrom}$ and $R_{\rm cn} = \SI{6}{\angstrom}$ for all the subsequent calculations with \textsc{gpumd}.

\subsection{Comparison between NEP-D3 and pure NEP for bilayer graphene}

\begin{figure}[ht]
\centering
\includegraphics[width=\columnwidth]{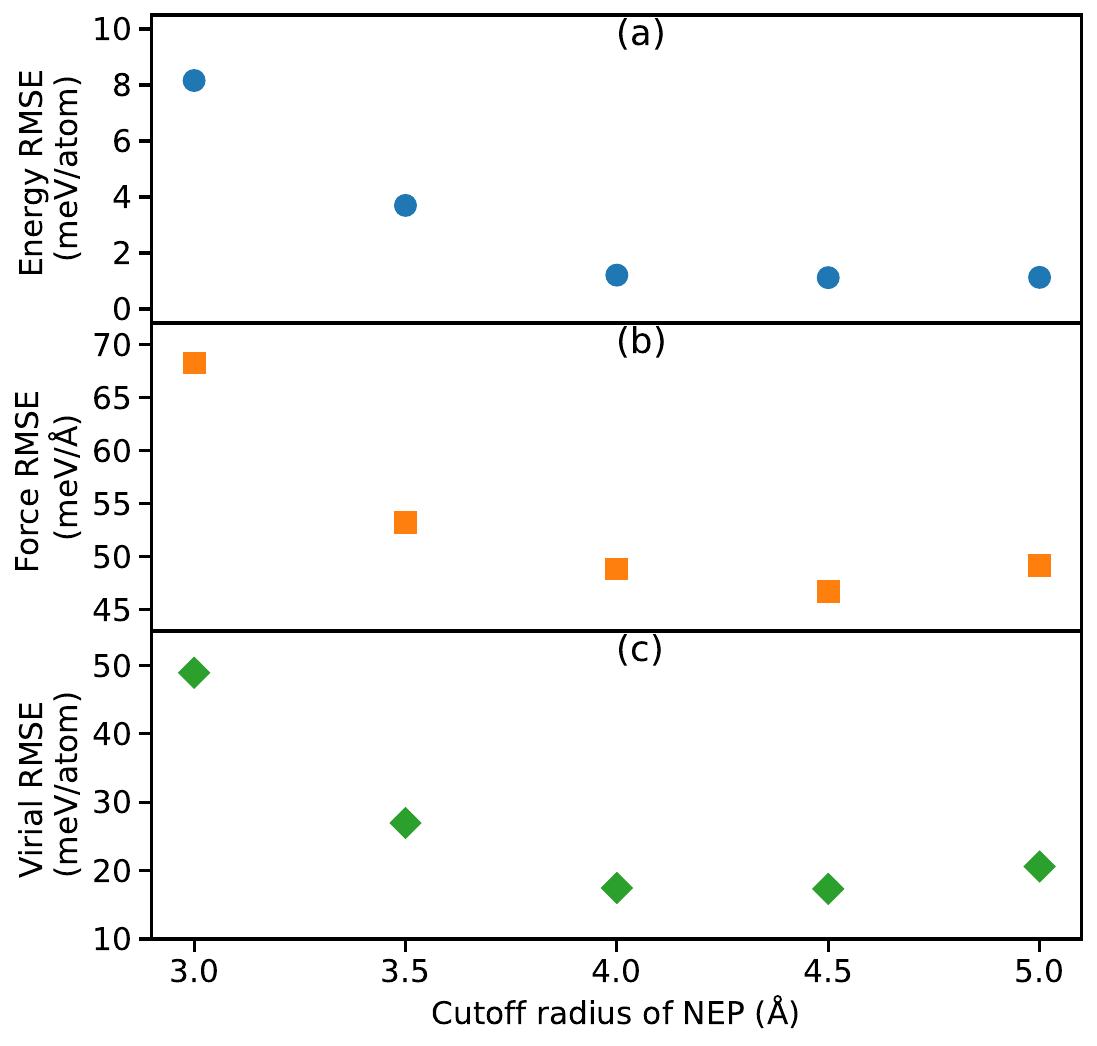}
\caption{\glspl{rmse} of (a) energy, (b) force, and (c) virial for \glspl{nep} with different cutoffs (with $r^{\rm R}_{\rm c}=r^{\rm A}_{\rm c}$) trained against the \gls{pbe} data set.}
\label{fig:rmse}
\end{figure}

We now compare the \gls{nep}-D3 and the pure-\gls{nep} approaches, taking bilayer graphene systems as an example. To this end, we generated a training data set consisting of bilayer graphene structures with different interlayer distances (from \SI{2}{\angstrom} to \SI{10}{\angstrom}) and relative lateral shifts (including the important AA, AB, and \gls{sp} stacking patterns and intermediate ones) as well as those from \gls{md} simulations (from \SI{300}{\kelvin} to \SI{1500}{\kelvin}) driven by the second-generation reactive empirical bond-order potential \cite{brenner2002second} in combination with the registry-dependent interlayer potential \cite{ouyang2018nanoserpents}. For all the structures, we performed single-point \gls{dft} calculations as details in Appendix \ref{sec:dft} to obtain energy, force, and virial data that are needed for \gls{nep} training. Two different kinds of reference data sets were considered: one based on the \gls{pbe} functional \cite{Perdew1996PRL} and the other based on \gls{pbe} combined with D3  \cite{grimme2011effect}. We labeled them as \gls{pbe} and \gls{pbe}-D3 data sets, respectively.

\begin{table}[thb]
\setlength{\tabcolsep}{0.75Mm}
\caption{The \glspl{rmse} for energy (\SI{}{\meV\per\atom}), force (\SI{}{\meV\per\angstrom}), and virial (\SI{}{\meV\per\atom}) of the \gls{nep} models versus the \gls{pbe}-D3 reference and the computational speeds (atom-step/second) in \gls{md} simulations of bilayer graphene with \num{40000} atoms using one GeForce RTX 4090 GPU card.}
\begin{tabular}{lllll}
\hline
\hline
\gls{nep} models   & Energy    & Force   & Virial  & Speed  \\
\hline
\gls{nep} (\SI{4.5}{\angstrom},  \SI{4.5}{\angstrom})-D3 & 1.16 & 46.72 & 15.38 & 6.29e6 \\
\gls{nep} (\SI{6.0}{\angstrom},  \SI{4.5}{\angstrom})    & 1.28 & 47.45 &  17.33 & 2.05e7 \\
\gls{nep} (\SI{8.0}{\angstrom},  \SI{4.5}{\angstrom})    & 1.25  & 48.08 &  16.09 & 1.53e7 \\
\gls{nep} (\SI{10.0}{\angstrom}, \SI{4.5}{\angstrom})    & 1.29  & 49.60 &  17.44 & 8.50e6 \\
\hline
\hline
\end{tabular}
\label{table:models}
\end{table}

\begin{figure*}[ht]
\centering
\includegraphics[width=1.8\columnwidth]{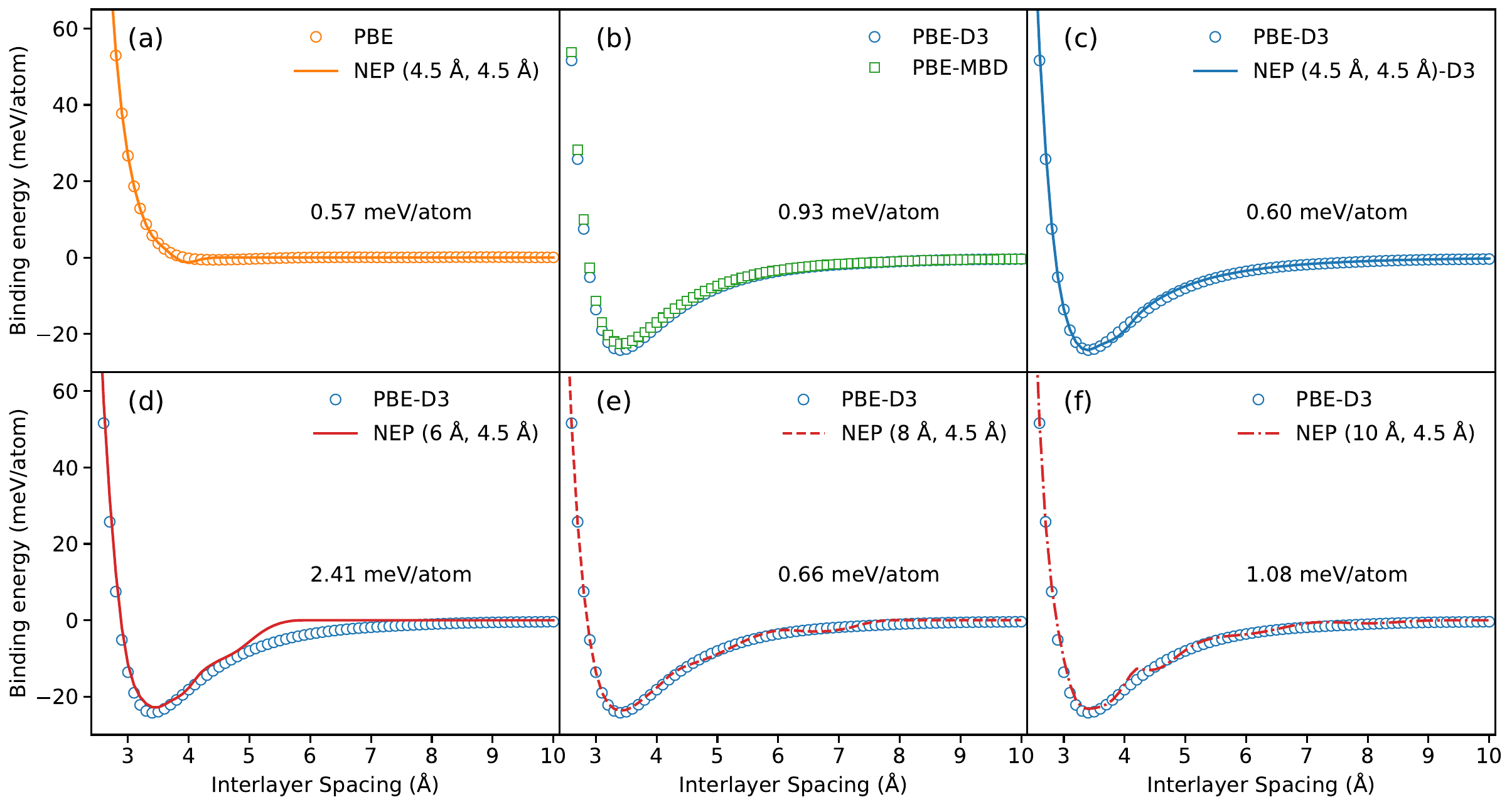}
\caption{Binding energy of AB-stacked bilayer graphene as a function of the interlayer distance computed from various approaches as indicated in each panel. The \gls{rmse} between the two curves in each panel is indicated. See text for details.}
\label{fig:binding}
\end{figure*}

To obtain a \gls{nep}-D3 model, we first trained a \gls{nep} model based on the \gls{pbe} data set. In this case, there is no need to use a large $r^{\rm R}_{\rm c}$ in \gls{nep} because the interactions from the \gls{pbe} functional are essentially short-ranged. We then took $r^{\rm R}_{\rm c}=r^{\rm A}_{\rm c}$ and tested the convergence of training accuracy with increasing cutoff. All the other hyper-parameters of \gls{nep} are listed in Appendix \ref{sec:nep_in}. \autoref{fig:rmse} shows that $r^{\rm R}_{\rm c}=r^{\rm A}_{\rm c}=\SI{4.5}{\angstrom}$ is quite optimal. By adding up the D3 part, we then obtained a compound potential model named as \gls{nep}(\SI{4.5}{\angstrom}, \SI{4.5}{\angstrom})-D3.
For the pure \gls{nep} models, we fixed the angular cutoff to \SI{4.5}{\angstrom} and considered three radial cutoffs: \SI{6}{\angstrom}, \SI{8}{\angstrom}, and \SI{10}{\angstrom}. This gives rise to three \gls{nep} models denoted as \gls{nep}(\SI{6}{\angstrom}, \SI{4.5}{\angstrom}), \gls{nep}(\SI{8}{\angstrom}, \SI{4.5}{\angstrom}), and \gls{nep}(\SI{10}{\angstrom}, \SI{4.5}{\angstrom}), respectively, which were trained against the \gls{pbe}-D3 data set.

The performances of the four models are compared in \autoref{table:models}. The \gls{nep}(\SI{4.5}{\angstrom}, \SI{4.5}{\angstrom})-D3 model has the highest accuracy in terms of energy, force, and virial. However, it has the lowest computational speed. The D3 part takes about 75\% of the computational cost in the \gls{nep}(\SI{4.5}{\angstrom}, \SI{4.5}{\angstrom})-D3 model. This reflects the high computational cost of D3 and also the high computational efficiency of \gls{nep}. Indeed, the \gls{nep} approach has been shown to be far more computationally efficient than other state-of-the-art \glspl{mlp} \cite{fan2022jcp}.
Because the dispersion interactions are weak forces, the \glspl{rmse} are not the best indicators for evaluating the accuracy. To better appreciate the higher accuracy brought about by the \gls{nep}-D3 approach compared to the pure \gls{nep} approach, we examine the binding and sliding energies in detail below.  

\autoref{fig:binding} shows the binding energy curves from the various calculation methods. The binding energies in bilayer graphene cannot be well captured by the \gls{pbe} functional, which is essentially zero at an interlayer distance of \SI{4.5}{\angstrom}, and this is the reason why a cutoff of \SI{4.5}{\angstrom} is sufficient for \gls{nep} to fit the \gls{pbe} data set (\autoref{fig:binding}(a)). By adding D3 to \gls{pbe}, the binding energies can be nicely captured and the results from D3 is very close to those from the \gls{mbd} \cite{Tkatchenko2012prl} (\autoref{fig:binding}(b)), which is one of the most accurate dispersion corrections presently available. By adding D3 to \gls{nep}(\SI{4.5}{\angstrom}, \SI{4.5}{\angstrom}), the resulting model, \gls{nep}(\SI{4.5}{\angstrom}, \SI{4.5}{\angstrom})-D3 is very close to \gls{pbe}-D3 (\autoref{fig:binding}(c)), and most of the errors come from the \gls{nep} part, with small extra errors from the truncation of the D3 part to $ R_{\rm pot} = \SI{12}{\angstrom}$ and $ R_{\rm cn} = \SI{6}{\angstrom}$. Without adding up D3, the pure \gls{nep} models with a relatively large radial cutoff can also partially capture the binding energies, but the curves are not as smooth as that from \gls{nep}-D3 (\autoref{fig:binding}(d)-(f)). 
The best results were obtained with a radial cutoff of \SI{8}{\angstrom}, which means that increasing the radial cutoff in \gls{nep} is not a feasible way to improve the accuracy regarding the \gls{vdw} interactions. To sum up, the \gls{nep}(\SI{4.5}{\angstrom}, \SI{4.5}{\angstrom})-D3 model outperform all the pure \gls{nep} models for describing the binding energies in bilayer graphene. It is expected that similar conclusions can be drawn for dispersion-dominated binding energies in layered materials.

\begin{figure}[ht]
\centering
\includegraphics[width=\columnwidth]{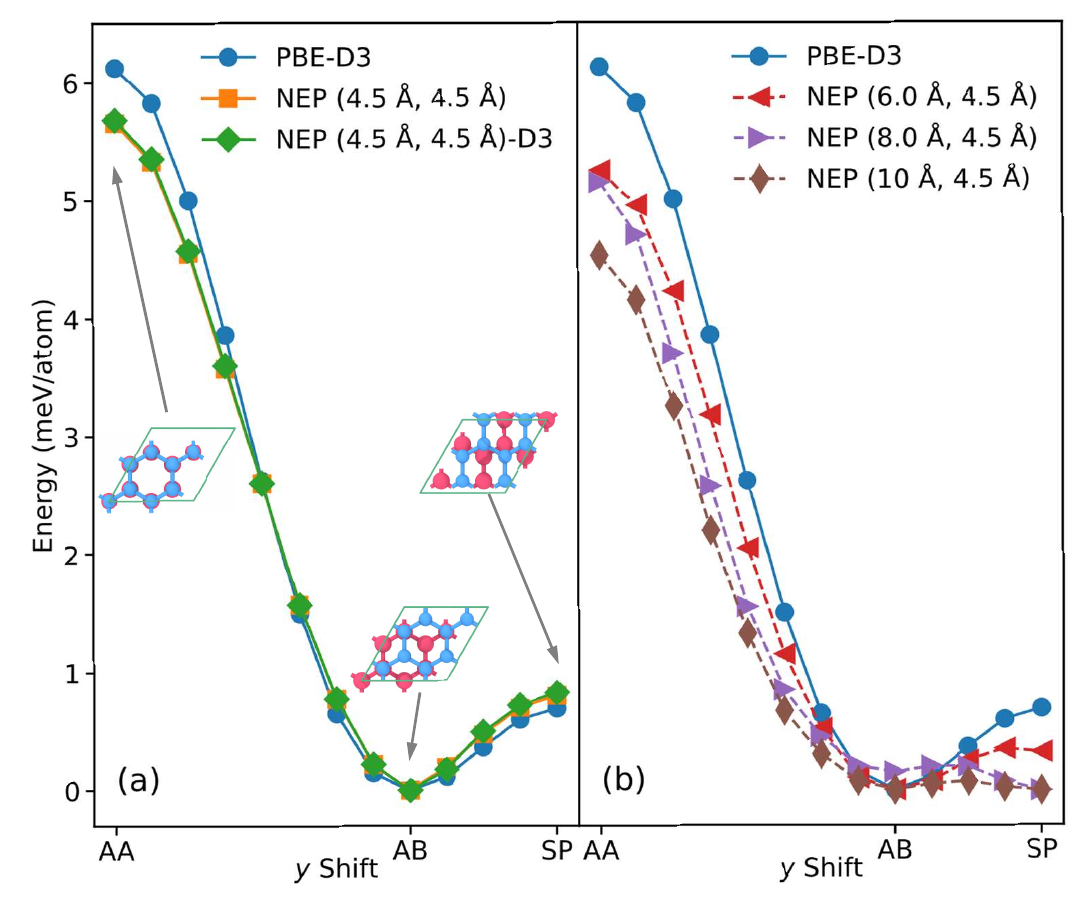}
\caption{Sliding energies along the AA-AB-SP path of bilayer graphene with a interlayer spacing of \SI{3.4}{\angstrom} computed from various approaches. Atoms from the top and bottom layers have different colors.}
\label{fig:sliding}
\end{figure}

In contrast to the binding energies, the sliding energies of bilayer graphene with equilibrium interlayer spacing (about \SI{3.4}{\angstrom}) are not dominated by the D3 part. Therefore, as shown in \autoref{fig:sliding}(a), results from \gls{nep}(\SI{4.5}{\angstrom}, \SI{4.5}{\angstrom}) are already very close to those from \gls{pbe}-D3 and adding D3 does not make significant changes. However, the resulting \gls{nep}(\SI{4.5}{\angstrom}, \SI{4.5}{\angstrom})-D3 model significantly outperforms the pure \gls{nep} models with larger radial cutoffs (\autoref{fig:sliding}(b)). One possible reason is that the long radial cutoff in a pure \gls{nep} model introduces extra features that are not needed for describing the sliding energies and thus complicates the training process. Using too large a cutoff than needed, the construction of a \gls{mlp} becomes more demanding since a large configuration space has to be explored and more descriptor components are needed to distinguish the atom environments \cite{Tokita2023jcp}. Indeed, among the three pure-\gls{nep} models, \gls{nep}(\SI{6}{\angstrom}, \SI{4.5}{\angstrom}) performs the best and \gls{nep}(\SI{10}{\angstrom}, \SI{4.5}{\angstrom}) the worst regarding the sliding energies (\autoref{fig:sliding}(b)).

\subsection{Applications to heat transport in MOFs}

After confirming the reliability of the combined \gls{nep}-D3 approach, we show its usefulness in practical \gls{md} simulations. In a previous work \cite{ying2023sub}, some of the present authors have studied heat transport in three typical \glspl{mof} (MOF-5, ZIF-8, and HKUST-1) using \gls{md} simulations with \gls{nep} models. The reference \gls{dft} data used for training these \gls{nep} models have no dispersion corrections. Because of the porous structures in \glspl{mof}, \gls{vdw} interactions between the organic chains are expected to have noticeable effects in the structural and dynamic properties. Here we study the effects of dispersion interactions on the thermal conductivity in the \glspl{mof}. 

\begin{figure}[ht]
\centering
\includegraphics[width=\columnwidth]{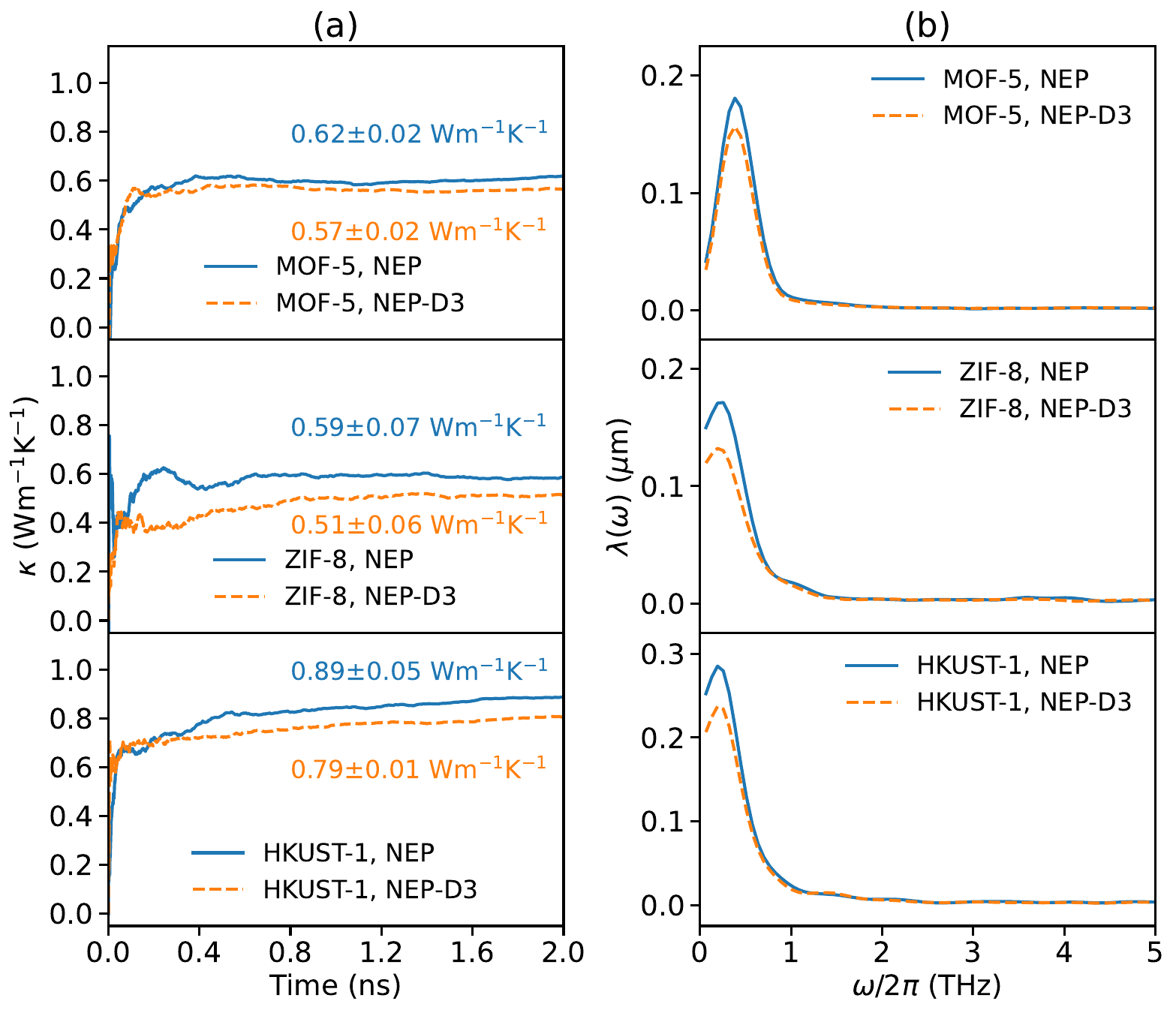}
\caption{(a) Thermal conductivity at \SI{300}{\kelvin} as a function of simulation time and (b) spectral phonon mean free path as a function of phonon frequency for MOF-5 (top), ZIF-8 (middle), and HKUST-1 (bottom). The converged thermal conductivity values with error estimates (from five independent runs) obtained from \gls{nep} and \gls{nep}-D3 are indicated.}
\label{fig:kappa}
\end{figure}

\autoref{fig:kappa} compares the thermal transport results from \gls{hnemd} simulations \cite{fan2019prb} at \SI{300}{\kelvin} using the previously constructed pure \gls{nep} models \cite{ying2023sub} and the combined \gls{nep}-D3 models obtained by adding up D3 ($R_{\rm pot}=\SI{12}{\angstrom}$, $R_{\rm cn}=\SI{6}{\angstrom}$). The \gls{md} simulation details are consistent with the previous work \cite{ying2023sub} and the relevant inputs are given in Appendix \ref{sec:run_in}.
For all the \glspl{mof}, the introduction of dispersion interactions consistently reduces the thermal conductivity, and the average amount of reduction being about 10\% (\autoref{fig:kappa}(a)). In the presence of dispersion interactions, the phonon mean free paths of the low-frequency phonons ($\omega/2\pi < 1$ THz) are reduced to some degree, but are still at the sub-micron scale (\autoref{fig:kappa}(b)). More importantly, the $\sim 10\%$ reduction of the thermal conductivity in MOF-5 still leaves the discrepancy between the calculated (\SI{0.57 \pm 0.02}{\watt\per\meter\per\kelvin}) and measured (\SI{0.32}{\watt\per\meter\per\kelvin}) results unresolved.

\section{Summary and conclusions}

In summary, we have made a GPU implementation of the D3 dispersion correction into the \textsc{gpumd} package and enabled its integration with \gls{nep} to form a combined \gls{nep}-D3 model. We demonstrated the superior accuracy of the \gls{nep}-D3 approach than the pure \gls{nep} approach by using bilayer graphene systems as an example, for which the dispersion interactions between the two layers play an important role for the binding energies. Although the dispersion interactions are not responsible for the sliding energies between the two layers, the presence of D3 in \gls{nep}-D3 allows for using a relatively short cutoff in the \gls{nep} part, which indirectly leads to better description of the sliding energies by the \gls{nep} part. The D3 dispersion correction we implemented can be readily added to available \gls{nep} models that have been trained against reference data without considering dispersion correction. As an example, we showed that adding D3 dispersion correction to the previous \gls{nep} models for \glspl{mof} \cite{ying2023sub}  leads to about 10\% reduced thermal conductivity. 

\begin{acknowledgments}
P. Ying was supported by the Israel Academy of Sciences and Humanities \& Council for Higher Education Excellence Fellowship Program for International Postdoctoral Researchers. Z. Fan was supported by National Natural Science Foundation of China (Project No. 11404033). 
\end{acknowledgments}

\vspace{0.5cm}
\noindent{\textbf{Data availability:}}

Complete input and output files for the \gls{nep} training of bilayer graphene and \glspl{mof} are freely available at \url{https://gitlab.com/brucefan1983/nep-data}.
The source code and documentation for \textsc{gpumd} are available
at \url{https://github.com/brucefan1983/GPUMD} and \url{https://gpumd.org}, respectively. 
The D3 dispersion correction is available starting from GPUMD-v3.9.

\vspace{0.5cm}
\noindent{\textbf{Declaration of competing interest:}}

The authors declare that they have no competing interests.

\appendix

\section{Details on the DFT calculations}
\label{sec:dft}

The \gls{dft} calculations were performed with the projector-augmented wave method \cite{Blchl1994PRB} implemented in \textsc{vasp} \cite{Kresse1996PRB, Kresse1999PRB}. We set a threshold of \SI{1e-7}{\eV} for the electronic self-consistent loop, with an energy cutoff of \SI{850}{\eV} for the plane-wave-basis set. We sampled the Brillouin zone using a $\Gamma$-centered grid with a k-point spacing of \SI{0.15}{\per\angstrom} and a Gaussian smearing with a width of \SI{0.1}{\eV}. The contents of the \verb"INCAR" input file for \textsc{vasp} are given below.

\begin{verbatim}
GGA      = PE       
LREAL    = Auto        
ENCUT    = 850    
IVDW     = 12 #remove this for PBE without D3
PREC     = Accurate 
KSPACING = 0.15     
KGAMMA   = .TRUE.   
ALGO     = Normal  
NSW      = 1           
IBRION   = -1
ISMEAR   = 0        
SIGMA    = 0.1     
EDIFF    = 1E-07    
NELM     = 150     
\end{verbatim} 

\section{Inputs for training the  NEP models}
\label{sec:nep_in}

The \gls{nep} models can be trained using the \verb"nep" executable in the \textsc{gpumd} package. The relevant hyperparameters are specified in the \verb"nep.in" input file. The contents of the \verb"nep.in" input file for training the \gls{nep}(\SI{4.5}{\angstrom}, \SI{4.5}{\angstrom}) model of bilayer graphene are given below. To train the \gls{nep} models with different radial cutoffs, one only needs to modify the parameters of the \verb"cutoff" keyword. 
\begin{verbatim}
type          1 C 
version       3        
cutoff        4.5 4.5
n_max         8 8      
basis_size    12 12   
l_max         4 2 0    
neuron        50       
lambda_1      0.05     
lambda_2      0.05     
lambda_e      1.0      
lambda_f      1.0      
lambda_v      0.1      
batch         10000   
population    50       
generation    300000     
\end{verbatim} 

\section{Inputs for thermal conductivity calculations}
\label{sec:run_in}

\gls{md} simulations with pure \gls{nep} models and \gls{nep}-D3 models can be performed by using the \verb"gpumd" executable in the \textsc{gpumd} package. The controlling parameters are specified in the \verb"run.in" input file. The contents of the \verb"run.in" input file for calculating the thermal conductivity using a \gls{nep}-D3 model are given below. To switch off the D3 contribution, one just needs to remove the line starting with the keyword \verb"dftd3".
\begin{verbatim}
potential       nep.txt
dftd3           pbe 12 6 
velocity        300

ensemble        npt_ber 300 300 100 0 10 1000
time_step       0.5
run             200000

ensemble        nvt_nhc 300 300 100
compute_hnemd   1000 0 0 2e-4
compute_shc     10 500 2 500 200
run             4000000
\end{verbatim}

\end{document}